\documentclass[authoryear]{elsarticle}

\journal{Annual Reviews in Control}

\usepackage{graphicx}

\usepackage{amssymb,amsmath}
\usepackage{enumitem}
\usepackage{mathtools}
\usepackage{comment}
\usepackage{url}
\usepackage[hyperindex,pdftex]{hyperref}

\usepackage{xcolor}

\usepackage{comment}
\usepackage{algorithm,algorithmic}
\usepackage{placeins}

\newdefinition{definition}{Definition}
\newdefinition{assumption}{Assumption}
\newdefinition{remark}{Remark}

\begin{document}

\begin{frontmatter}



\title{An Optimal Predictive Control Strategy for COVID-19 \\ (SARS-CoV-2) Social Distancing Policies in Brazil}

\author[adufsc]{Marcelo M. Morato}
\ead{marcelomnzm@gmail.com}
\author[adFACE]{Saulo B. Bastos}
\author[adFACE,adINCT,adunb]{Daniel O. Cajueiro}
\author[adufsc]{Julio E. Normey-Rico}


\address[adufsc]{Renewable Energy Research Group (\emph{GPER}), Departamento
  de Automa\c{c}\~ao e Sistemas (\emph{DAS}), \\Universidade Federal de Santa Catarina,
  Florian\'opolis, Brazil.}
  
  \address[adFACE]{Departamento de Economia, FACE, Universidade de Bras\'{i}lia (UnB), Campus Universit\'{a}rio Darcy Ribeiro, 70910-900, Bras\'{i}lia, Brazil.}

  \address[adINCT]{Nacional Institute of Science and Technology for Complex Systems (INCT-SC).}

\address[adunb]{LAMFO, FACE - Universidade de Bras\'{i}lia (UnB), Campus Universit\'{a}rio Darcy Ribeiro, 70910-900, Bras\'{i}lia, Brazil.}







\begin{abstract}
The global COVID-19 pandemic (SARS-CoV-2 virus) is the defining health crisis of our century.  Due to the absence of vaccines and drugs that can help to fight it, the world solution to control the spread has been to consider public social distance measures that avoids the saturation of the health system. In this context, we investigate a Model Predictive Control (MPC) framework to determine the time and duration of social distancing policies. We use Brazilian data in the period from March to May of 2020. The available data regarding the number of infected individuals and deaths suffers from sub-notification due to the absence of mass tests and the relevant presence of the asymptomatic individuals. We estimate variations of the SIR model using an uncertainty-weighted Least-Squares criterion that considers both nominal and inconsistent-data conditions. Moreover, we add to our versions of the SIR model an additional dynamic state variable to mimic the response of the population to the social distancing policies determined by the government that affects the speed of COVID-19 transmission. Our control framework is within a mixed-logical formalism, since the decision variable is forcefully binary (the existence or the absence of social distance policy). A dwell-time constraint is included to avoid harsh shifting between these two states. Finally, we present simulation results to illustrate how such optimal control policy would operate. These results point out that no social distancing should be relaxed before mid August 2020. If relaxations are necessary, they should not be performed before the beginning this date and should be in small periods, no longer than $25$ days. This paradigm would proceed roughly until January/2021. The second peak of infections, which has a forecast to the beginning of October, can be reduced if the periods of no-isolation days are shortened.
\end{abstract}

\begin{keyword}
COVID-19 \sep Model Predictive Control \sep On-Off Control \sep Social Distancing \sep Brazil.
\end{keyword}

\end{frontmatter}


\section{Introduction}

The COVID-19 pandemic seems to be the global health crisis of our time. Scientist first identified this virus (SARS-CoV-2) in humans in Wuhan, in the province of Hubei, China by December 2019. It causes severe acute respiratory syndrome which can become potentially fatal. The WHO estimated by the end of March that the number confirmed cases was reaching the order $70000$, with more than $33000$ confirmed deaths. Now, by the end of April, it has already spread  to almost every country of the world, infecting 3,019,246 and killing more than 208,112 people \cite{who20200425}. Its spread is rapid and efficient and it seems that to tackle this pandemic, global scientific efforts are necessary \cite{bedford2019new}. Since vaccines have not yet been developed, most countries have adopted measures to ensure social distancing, aiming to avoid the spread \cite{Adam2020}. It seems that the COVID-19 has posed an unique question regarding what are the viable public policies necessary to handle its spread.

\begin{figure}[htb]
	\centering
		\includegraphics[width=\linewidth]{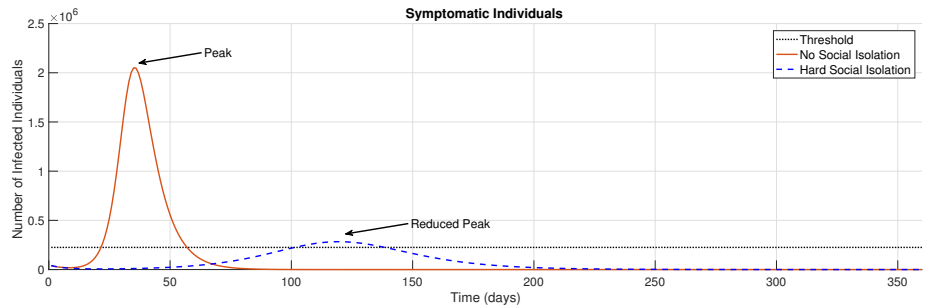}%
	\caption{Necessity of Social Isolation}
	\label{IsolationNeeded}
\end{figure}

The idea behind social distance is to prevent health systems from becoming saturated due to large amounts of COVID-19 patients being treated at the same time. Therefore, with social distancing policies, the health systems do not have to deal with hospital bed shortages associated with a large peak of infections, since the demands for treatment become distributed over time. Figure \ref{IsolationNeeded} illustrates the evolution of symptomatic individuals due to the SARS-CoV-2 virus with respect to no isolation and hard social isolation policies. The threshold represents an estimate for the number of available Intense Care Unit (ICU) hospital beds.

\FloatBarrier

However, social distancing measures exhibit at least three ambiguous side-effects, which policy makers should take into account:
\begin{enumerate}
    \item If the governments interrupt the social distancing policy before the correct time, the measure is only able to shift the pattern of contamination to the future, which does not help diminishing the problem of saturating the health system due to excessive demand for ICU beds. This topic has been throughly discussed by \cite{hellewell2020feasibility};
    \item Countries that implement rigid social distancing measures have seen devastating economic effects. The recent papers by \cite{NBERw26882,Gormsen2020} elaborate on this issue;
    \item A large part of the population may not be immunized and might suffer from future waves of COVID-19 infections, after the social distancing measures take place. Furthermore, recurrent wintertime outbreaks of SARS-CoV-2 virus will probably occur after this initial pandemic dissemination \cite{Kissler2020}.
\end{enumerate} 

Therefore, it becomes of fundamental importance to predict the correct time and the duration of social distancing interventions. Well-designed social distance policies may help to control the evolution of the disease, to avoid the saturation of the heath systems and to minimize the economic side effects caused by them.  

In this paper, the Brazilian context is taken into account \cite{werneck2020covid}. Brazil is a continent-sized tropical\footnote{Recent research point out that high temperatures may favor the spread of this virus; see the work by \cite{auler2020evidence}.} country and it has already been facing many issues due to the COVID-19 pandemic. The country has $26$ federated states, which have been choosing different social distancing measures since mid-March\footnote{Throughout this paper, the Year/Month/Day notation is used.}. The federal government is reluctant to implement nation-wide policies, disclaiming that the negative economic effects are too steep and that social distancing is an erroneous choice \cite{THELANCET20201461}; the government suggests that the economy cannot stop and that herd immunity could be a solution to this pandemic. The expected impacts of the disease in Brazil are catastrophic \cite{ismaeluniversal,rocha2020expected}. Moreover, due to lack of testing, Brazil is only accounting for patients with severe symptoms or those how have died; therefore, a huge percentage of sub-notification (over $90 \, \%$) has been reported \cite{silva2020bayesian, rocha2020expected, delatorre2020tracking}. The daily reports ("measurements") delivered by the Ministry of Health, collecting number of infected and deceased patients, only gives an impression of the virus spread of past moment, since, in average, a person will exhibit acute symptoms only $20$ days of the infection.

The first death due to the SARS-CoV-2 virus in Brazil was registered in March $17$, while the first case was officially notified in February $26$. Nonetheless, recent papers \cite{delatorre2020tracking, rodriguez2020covid} point out that the virus was already present in Brazil since the end of January, before the Carnival festivals. Through inferential statistics, \cite{delatorre2020tracking} acknowledge that community transmission has been ongoing in the state of S\~ao Paulo since the beginning of February (over one month before the official reports).

\begin{figure}[htb]
	\centering
		\includegraphics[width=\linewidth]{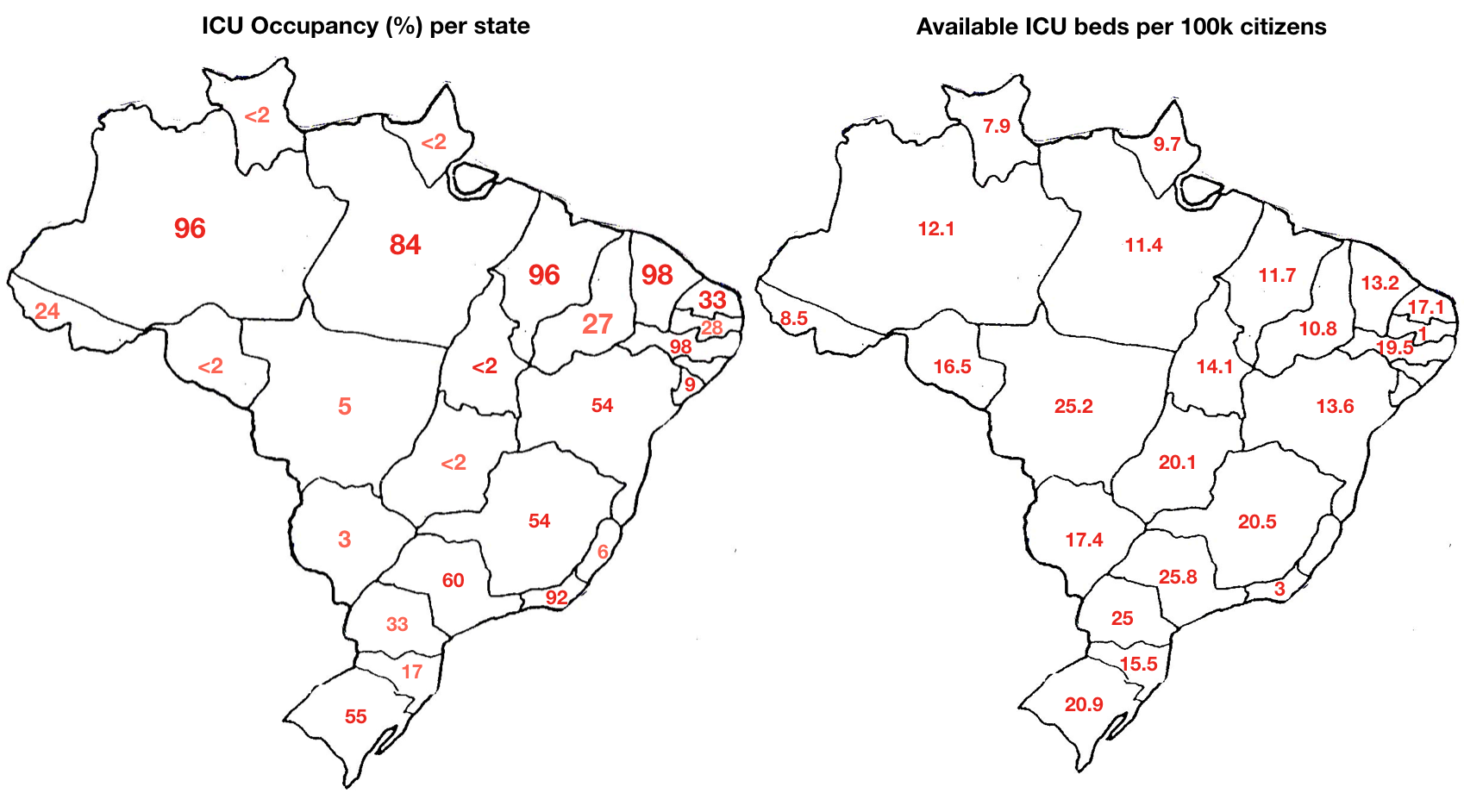}
	\caption{ICU Beds and occupancy rate in Brazil, per state ($20/04/30$).}
	\label{ICUbedsBr}
\end{figure}

Even though a strong public health system is available in Brazil, as of April $30$, many states were already exhibiting a near-collapse situation, with over $95\, \%$ of ICU beds occupied with COVID-19 patients. This is illustrated in Figure \ref{ICUbedsBr}, which show the ICU occupancy rate and the number of available in many states \cite{CFM2020}. Clearly, the situation is already border-lining.

\FloatBarrier

Motivated by the previous discussion, in this paper we investigate the problem of controlling the evolution of the COVID-19 pandemic in Brazil using optimal social distancing policies, which are designed through a Model Predictive Control (MPC) framework.

The MPC framework is a widespread optimal control method for the control of processes subject to constraints \cite{camacho2013model}. MPC allows to explicitly consider the effect of input, output and state
constraints in the control design procedure, which is rather convenient. As any standard discrete control method \cite{skogestad2007multivariable}, MPC-formulated laws stand for piece-wise-constant signals for sampled-data systems, which  is clearly the scope of the COVID-19 dissemination process, since it is measured daily through the number of infected and deceased individuals.

The frameworks to predict and control the complex dynamics of the SARS-CoV-2 virus spread are definitely  mixed discrete-continuous problems under multiple objectives due to the nature of the problem (daily measurements and piece-wise constant control). MPC certainly fits this context.

We must remark that many research papers have demonstrated the application and validation of this control tool for health-related, biological and ecological regulation purposes \cite{ionescu2008robust, zurakowski2004hiv, moscoso2016intra, de2019model}. Thus, MPC fits naturally to COVID-19 social distancing control problem. Any possible COVID-19 control framework should take into account social distancing constraints and goal of minimizing the peak of infected individual, which is viable through the MPC paradigm. The main contributions of this paper are the following:
\begin{itemize}
    \item We present two modified versions of the Susceptible-Infected-Recovered (SIR) model \cite{Kermack1927}, embedding the effects of social distancing measures in the evolution of the disease;
    \item We propose an additional dynamic state variable, which models the response of the population to social distancing measures enforced by the government. We also use this state variable to forecast the reduction of the speed of transmission of the virus, with respect to enacted distancing policy;
    \item Due to the fact that large error margins have been reported regarding the available COVID-19 data and statistics in Brazil (see  \cite{ImperialCollegeNew}), we perform an uncertainty-weighted Least-Squares criterion to estimate the parameters of the virus infection/spread model, considering both nominal and inconsistent-data conditions.
    \item Based on these uncertain models, we propose an MPC-based control framework to determine in real time whether to apply or not the social distancing policy. This control strategy resides in the solution of a Mixed-Integer Dynamic Programming Problem, at each sampling instant (day), according to new available datasets (number of infected and deaths). The constraints of the MPC procedure are given with respect to the number of available ICU hospital beds in the country. The MPC also accounts for a minimal dwell-time on each control action (no isolation, complete lock-down), so that frequent social distancing policy shifting does not happen.
\end{itemize}

Note that, for the sake of practical purposes, we estimate model parameters and distancing policies using real data from the Brazilian Ministry of Health, in a fashion similar to the estimation scheme presented by \cite{bastos2020}. 

Our paper relates to some other recent papers that investigate the COVID-19 pandemic from a control viewpoint. There are some recent works that have also inserted a control variable in the available epidemiological model in order to emulate the control of infections using strategies such as vaccination, isolation, culling and self-isolation \cite{Bolzoni2017, Piunovskiy2019,piguillem2020optimal}. In particular, we state that the recent paper \cite{piguillem2020optimal} has also addressed the issue of optimal social distancing COVID-19 policies. A case study regarding Belgium has been presented in \cite{alleman2020covid}, considering a continuous MPC policy; a robust MPC method, regarding the COVID-19 spread in Germany, has been investigated by \cite{kohler2020robust}.

We must stress that this paper differs from \cite{piguillem2020optimal,alleman2020covid} and \cite{kohler2020robust} in three main points:
\begin{enumerate}[label =(\roman*)]
    \item We consider the possibility of large amounts of uncertainty regarding the infected/deaths measurements (which stands true for the Brazilian case, which is ours). The previous papers considered relatively small parametric uncertainties added to the model (identified with real data). In our case, we consider large amounts of uncertainty and use uncertainty-embedded models derived through a series of identification runs.
    \item As in \cite{alleman2020covid, kohler2020robust}, we design and synthetize an optimal control strategy with a Model Predictive Control formalism, that holds recursive feasibility and stabilization properties and has wide industrial practice. Anyhow, the MPC design in this paper follows a mixed-integer approach and embeds a dwell-time constraint, which had not been tested in the previous paper. Furthermore, the chosen control input differs: in \cite{alleman2020covid}, the control input is the actual isolation parameter, while in \cite{kohler2020robust} it directly affects the infection and transmission rates. In this paper, the control input stands for a new model variable which indicates the government enacted social distancing policy, which affects the average isolation observed in the population, that then meddles with the transmission and infection dynamics.
     \item We model the response of the population to these social isolation rules with an additional dynamic variable (and fit our variations of the SIR models accordingly).
    \item The considered dwell-time constraint ensures that the social distancing policy remains constant for at least $N_m$ days, avoiding frequent shifting between isolation and non-isolation states (which obviously would cause ambiguity and confusion to the population). 
\end{enumerate}

This work is organized as follows: In Section \ref{sec:models}, we introduce the SIR models used in this work and the modifications we make in order to model people's response to the social distancing policies. In Section \ref{sec:estimation}, we present the approach used to estimate the parameters of the models; therein, we also present the model-data fitting results. Section \ref{sec:control} presents the optimal control scheme used to harness the evolution of the disease along time. Finally, Section \ref{sec:discussion} presents the main conclusions of the work.

\section{Epidemiological models}
\label{sec:models}

Recent literature \cite{peng2020epidemic,kucharski2020early} shows that the infection rate and evolution dynamics of the SARS-CoV-2 virus can be adequately described by Susceptible-Infected-Recovered (SIR) models, which were originally presented by \cite{Kermack1927}.

In this Section, we present two modified versions of the SIR model that take into account the effects of social distancing measures and embedded them to the evolution dynamic of the disease. The new variables account for the dynamics of the population response to such social distancing measures (enacted by local governments). 

\subsection{Original SIR Model and Modifications Regarding COVID-19}

The SIR describes the spread of a given disease with respect to a population split into three non-intersecting classes, which stand for:
\begin{itemize}
    \item Susceptible individuals (S), who are prone to contract the disease;
    \item Infected individuals (I), which currently have the disease;
    \item Recovered individuals (R), who have already recovered from the disease. 
\end{itemize}   
Due to the evolution of the spread of the disease, the size of each of these classes change over time and the total population size \(N\) is the sum of these three classes, as follows:
\begin{equation}
N(t)=S(t)+I(t)+R(t)\label{eq:Nconstant} \, \text{.} 
\end{equation}

In the SIR model, the parameter \(\beta\) stands for the average number of contacts that are sufficient for transmission of the virus from one individual, per unit of time \(t\). Therefore,  \(\beta I(t)/N(t)\) determines the average number of contacts that are sufficient for transmission from infected individuals, per unit of time, to one susceptible individual; and \((\beta I(t)/N(t))S(t)\) determines the number of new cases per unit of time due to the amount of \(S(t)\) susceptible individuals (they are ``available for infection''). 

Furthermore, the parameter \(\gamma\) stands for the recovery rate, which is the rate that each infected individual recovers (or dies). This parameter characterizes the amount of individuals that ``leaves'' the infected class, considering a constant probability quota per unit of time.

Based on these definitions, the SIR dynamics are:
\begin{equation}
    \begin{array}{rcl}
    \displaystyle \frac{dS}{dt}(t) & = & \displaystyle -\frac{\beta I(t)S(t)}{N(t)}\\[3mm]
    \displaystyle \frac{dI}{dt}(t) & = & \displaystyle \frac{\beta I(t)S(t)}{N(t)} - \gamma I(t) \\[3mm]
    \displaystyle \frac{dR}{dt}(t) & = & \displaystyle \gamma I(t)  
    \end{array}\;\;\;\textrm{\bf [SIR]}. 
\label{eq:SIR}\end{equation}


Since the SIR model is used herein to describe a short-term pandemic outbreak, we do not consider the effects of demographic variations. Despite recent discussion regarding the possibilities of reinfection\cite{del2020covid}, we assume that the recovered individuals will not be reinfected (at least for simplicity purposes), i.e. an individual does not contract the disease twice. We will not implement this first SIR model, it is only included for the sake of referencing.

Indeed, the model should also include the dynamic relationships that appear due to the fraction of people that unfortunately die from the disease. Thus, we include a parameter \(\rho\), which stands for the probability of an individual form the infected  class \(I(t)\) dying from infection before recovering, as suggested in \cite{keeling2011}. In this case, the following set of Equations arise:
\begin{equation}
    \begin{array}{rcl}
    \displaystyle \frac{dS}{dt}(t) & = & \displaystyle - \frac{\beta I(t) S(t)}{N(t)}\\[3mm]
    \displaystyle \frac{dI}{dt}(t) & = & \displaystyle \frac{\beta I(t) S(t)}{N(t)} - \gamma I(t)  - \frac{\rho}{1 - \rho} \gamma I(t) = \frac{\beta I(t) S(t)}{N(t)} - \frac{\gamma I(t)}{1 - \rho}\\[3mm]
    \displaystyle \frac{dR}{dt}(t) & = & \displaystyle \gamma I(t) \\[3mm]
    \displaystyle \frac{dD}{dt}(t) & = & \displaystyle \frac{\rho}{1-\rho} \gamma I(t)\\
    \end{array}\;\;\;\textrm{\bf [SIRD]},
\label{eq:SIRD}
\end{equation}
\noindent where \(\frac{\rho}{1-\rho} \gamma I(t)\) stands for the number of people from the population that die due to the disease, per unity of time; and \(D(t)\) is the number of people that die due to the disease. Note that, in this case, the number of individuals in the population reduces due to the infection according to $$\frac{dN}{dt}(t) = -\frac{\rho}{1 - \rho} \gamma I(t) \, \text{.}$$ For the ease of reference, this adaptation of the SIR model is named hereafter as the ``{\bf SIRD}'' (Susceptible-Infected-Recovered-Dead) model. 

Since, in the case of the SARS-CoV-2 virus, there is a relevant percentage of the infected individuals that are asymptomatic, we split the class of infected individuals into the classes of symptomatic and asymptomatic individuals, as suggested in \cite{Robinson2013,Arino2008,Longini2004}: 

    \begin{equation}
    \begin{array}{rcl}
    \displaystyle \frac{dS}{dt}(t) & = & \displaystyle  - (\beta_A I_A(t) + \beta_S I_S(t))\frac{ S(t)}{N(t)} \\[3mm]
    \displaystyle \frac{dI_A(t)}{dt} & = & \displaystyle  (1-p)(\beta_A I_A(t) + \beta_S I_S(t))\frac{ S(t)}{N(t)} - (\gamma_A) I_A(t) \\[3mm]
    \displaystyle \frac{dI_S}{dt}(t) & = & p (\beta_A I_A(t) + \beta_S I_S(t))\frac{ S(t)}{N(t)} - \frac{\gamma_S I_S(t)}{1 - \rho} \\[3mm]
    \displaystyle \frac{dR_A}{dt}(t) & = & \displaystyle \gamma_A I_A(t) \\[3mm]
    \displaystyle \frac{dR_S}{dt}(t) & = & \displaystyle \gamma_S I_S(t) \\[3mm]
    \displaystyle \frac{dD}{dt}(t) & = & \displaystyle \frac{\rho}{1 - \rho} \gamma_S I_S(t) \\[3mm]
    \end{array}\;\;\;\textrm{{\bf [SIRASD]}},
    \label{eq:IAISwithP} 
    \end{equation}
\noindent where $I_A$ ($R_A$) is the number of asymptomatic infected (recovered) individuals, $I_S$ ($R_S$) is the number of symptomatic infected (recovered) individuals and $p$ is the proportion of individuals who develop symptoms. For ease of reference, this latter model is named hereafter as the ``{\bf SIRASD}'' (Susceptible-Infected-Recovered-Asymptomatic-Symptomatic-Dead) model. The SIRASD model has been widely used in the recent literature to describe the COVID-19 pandemics \cite{piguillem2020optimal}. Just like with SIRD model, the original condition that $N(t)$ is constant over time form the SIR model no longer holds. Therefore, to evaluate the variation of the population size $N(t)$ over time, one needs to integrate $$\frac{dN}{dt}(t) = -\frac{\rho}{1 - \rho} \gamma_S I_S(t) \, \text{.}$$

\subsection{The New Control-Embedded Models}

In order to design and synthesize effective control strategies for social distancing (public) policies, to be oriented to the population by local governments, these previous SIRD and SIRASD models must be adapted to include the dynamics and effects of social distancing.

Therefore, a new differential equation is proposed to the response of population to such social distancing rules. These new dynamic is appended to those of $S(t)$, $I(t)$ and $D(t)$, which also become affected by the amount of social distancing at a present time. 

We will denote \(\psi(t)\) as this varying parameter which account for the population response: $\psi = 0$ stands for the case of a complete lock-down (in a hypothetical and unachievable sense that there would not be any more contact between individuals in the population), while $\psi = 1$ stands for a no-isolation state, where the population is behaving ``as usual'', with regular activities as before the pandemic. The differential equation which models the evolution of $\psi (t)$ is the following:

\begin{equation}
    \begin{array}{rcl}
    \displaystyle \frac{d\psi}{dt}(t) & = & \alpha_{\mathrm{Off}}\left(1-\psi(t)\right)\left(1-u(t)\right) + \alpha_{\mathrm{On}}\left(K_\psi(t)\psi_{\mathrm{inf}}-\psi(t)\right)u(t)
    \end{array}\;\;\;\textrm{\bf [C]}.  
\label{eq:SIRDC}
\end{equation}
where $u(t)$ is a binary variable which determines the social isolation policy regulated by the government (this signal will be later on determined by the proposed optimal controller): for the cases when social isolation is determined, an "On" state is set, with $u = 1$; when the government does not enact an isolation measure, an "Off" state is set, with $u =0$. Note that $\alpha_{\mathrm{On}}$ and $\alpha_{\mathrm{Off}}$ are settling-time parameters which relate to the average time the population takes to respond to the enacted social isolation measures.

The dynamic equation above [C] accounts for these two "On"/"Off" possibilities: 
\begin{itemize}
\item ``Off'') if the government determines no isolation is necessary, it follows that $u=0$ and $\frac{d\psi}{dt}(t) = \alpha_{\mathrm{Off}}(1-\psi(t))$, meaning that roughly after $\frac{5}{\alpha_{\mathrm{Off}}}$ days, $\psi$ converges\footnote{Note that $\lim_{t\to \infty} \left(\frac{d\psi}{dt}(t) = \alpha_{\mathrm{Off}}(1-\psi(t))\right) \to \left(\psi \to 1\right)$.} to $1$;
\item ``On'') if the government determines that a hard isolation is needed, it follows that $u=1$ and $\frac{d\psi}{dt}(t) = \alpha_{\mathrm{On}}(K_\psi(t)\psi_{\mathrm{inf}}-\psi(t))$, meaning that roughly after $\frac{5}{\alpha_{\mathrm{On}}}$ days, $\psi$ converges\footnote{Note that $\lim_{t\to \infty} \left(\frac{d\psi}{dt}(t) = \alpha_{\mathrm{On}}(K_\psi(t)\psi_{\mathrm{inf}}-\psi(t))\right) \to \left(\psi \to \overline{K_\psi}\psi_{\mathrm{inf}}\right)$.} to $\overline{K_\psi}\psi_{\mathrm{inf}}$, where $\psi_{\mathrm{inf}}$ stands for the "hardest" isolation observed in practice (which is usually $> 0$); $\overline{K_\psi}$ is a static gain. In practice, [C] includes a time-varying gain $K_\psi(t)$ which holds the limit property $\lim_{t\to \frac{5}{\alpha_{\mathrm{On}}}} K_\psi (t) \to \overline{K_\psi}$. This term is included to represent the relationship between the actual observed isolation $\psi$ and the "hardest" isolation $\psi_{\mathrm{inf}}$. 
\end{itemize}

\begin{remark}
Since [C] represents a first-order differential system, the settling-time constants $\alpha_{\mathrm{On}}$ and $\alpha_{\mathrm{Off}}$ determine the convergence speed of $\psi(t)$. Notice that, for an arbitrary system $\frac{dx}{dt}(t) = \alpha_x\left(x_f-x(t)\right)$, $x(t)$ reaches $0.99x_f$ in $5/\alpha_x$ units of time, since the solution for this differential equation is $x(t) = \left(1-e^{-\alpha_x t}\right)x_f$, with $\alpha_x > 0$, for which $x(5/\alpha_x) = 0.99x_f$.
\end{remark}

Since $\psi(t)$ is the average people's response to public policies to reduce the spread of the virus (such as isolation measures or incentive to wear masks), it affects the transmission factors and, thus, we replace \(\beta\) in the SIRD model by \(\psi(t) \beta\) and \(\beta_A\) and \(\beta_S\) in the SIRASD model, respectively, by \(\psi(t) \beta_A\) and \(\psi(t) \beta_S\). It is worth mentioning that the population response to the On-Off isolation policy control may depend on factors such as the incremental number of deaths and the amount of information they known about the disease. Anyhow, since this factor depends on people's choices, and thus becomes rather difficult to quantify, we opt for the simplicity of the previous differential equation [C], since we may partially estimate its parameters with the available data.

Finally, we include these new dynamics into the SIRD and SIRASD models as discussed. In this case, we get the following models that we denote {\bf SIRDC} (Susceptible-Infected-Dead-with Control) model:
\begin{equation}
    \begin{array}{rcl}
    \displaystyle \frac{dS}{dt}(t) & = & \displaystyle - \frac{\psi(t)\beta I(t) S(t)}{N(t)}\\[2mm]
    \displaystyle \frac{dI}{dt}(t) & = & \displaystyle \frac{\psi(t)\beta I(t) S(t)}{N(t)} - \frac{\gamma I(t)}{1 - \rho}\\[2mm]
    \displaystyle \frac{dR}{dt}(t) & = & \displaystyle \gamma I(t) \\[2mm]
    \displaystyle \frac{dD}{dt}(t) & = & \displaystyle \frac{\rho}{1-\rho} \gamma I(t)\\[2mm]
    \displaystyle \frac{d\psi}{dt}(t) & = & \displaystyle \alpha_{\mathrm{Off}}\left(1-\psi(t)\right)\left(1-u(t)\right) + \alpha_{\mathrm{On}}\left(K_\psi(t)\psi_{\mathrm{inf}}-\psi(t)\right)u(t)\\ K_\psi(t) & = & \displaystyle 1-\gamma_K\gamma \frac{\rho}{1-\rho}\frac{I(t)}{N(t)} \, \text{.}
    \end{array}\;\;\;\textrm{\bf [SIRDC]};   
\label{eq:SIRDC}
\end{equation}

\noindent and {\bf SIRASDC} (Susceptible-Infected-Asymptomatic-Symptomatic-Dead-with Control):
\begin{equation}
\begin{array}{rcl}
\displaystyle \frac{dS}{dt}(t) & = & \displaystyle  - \psi(t)(\beta_A I_A(t) + \beta_S I_S(t))\frac{ S(t)}{N(t)} \\[2mm]
\displaystyle \frac{dI_A}{dt}(t) & = & \displaystyle  (1-p)\psi(t)(\beta_A I_A(t) + \beta_S I_S(t))\frac{ S(t)}{N(t)} - (\gamma_A) I_A(t) \\[2mm]
\displaystyle \frac{dI_S}{dt}(t) & = & \displaystyle p \psi(t)(\beta_A I_A(t) + \beta_S I_S(t))\frac{ S(t)}{N(t)} - \frac{\gamma_S I_S(t)}{1 - \rho} \\[2mm]
\displaystyle \frac{dR_A}{dt}(t) & = & \displaystyle \gamma_A I_A(t) \\[2mm]
\displaystyle \frac{dR_S}{dt}(t) & = & \displaystyle \gamma_S I_S(t) \\[2mm]
\displaystyle \frac{dD}{dt}(t) & = & \displaystyle \frac{\rho}{1 - \rho} \gamma_S I_S(t) \\[2mm]
 \displaystyle \frac{d\psi}{dt}(t) & = & \alpha_{\mathrm{Off}}\left(1-\psi(t)\right)\left(1-u(t)\right) + \alpha_{\mathrm{On}}\left(K_\psi(t)\psi_{\mathrm{inf}}-\psi(t)\right)u(t)\\ K_\psi(t) &=& \displaystyle 1-\gamma_K\gamma_A\frac{\rho}{1-\rho}\frac{I_A(t)}{N(t)}
\end{array}\;\;\;\textrm{{\bf [SIRASDC]}},
\label{eq:SIRASDC} 
\end{equation}

\noindent which will be used for identification and control purposes.

\section{Tuning of the SIRDC / SIRASDC Models \label{sec:estimation}}

In this Section, we present the results concerning the estimation of the epidemiological parameters of Eq. (\ref{eq:SIRASDC}).

We must, at first, re-affirm that recent literature regarding the Brazilian COVID-19 context has raised attention to the large amount of sub-notification in the country \cite{ImperialCollegeNew,THELANCET20201461, silva2020bayesian,rocha2020expected,rodriguez2020covid}. Therefore, we progress by embedding the uncertainty regarding the available datasets to the identification procedure.

The following identification is based the real data provided by the Ministry of Health of Brazil from February 25, 2020 to May 8, 2020. Specifically, we consider the cumulative number of infected individuals ($Z(t) = I(t) + R(t) + D(t)$) and the number of deaths $D(t)$.

To incorporate the issue of sub-notification, we assume that the data provided by the Ministry of Health is corrupted. Instead of using the original data, we assume: 
\begin{eqnarray}
\label{Dtuncertain}
D^{\text{nominal}}(t) &=& q_D \, D(t) \, \text{,} \\ \label{Ztuncertain}
Z^{\text{nominal}}(t) - D(t) &=& q_I \, (Z(t) - D(t)) \, \text{,}
\end{eqnarray}
where the instances of $Z^{\text{nominal}}(t)$ and $D^{\text{nominal}}(t)$ represent the data provided by Ministry of Health. The parameters $q_I \in [0,1]$ and $q_D \in [0,1]$ are uncertainty measures that provide a relationship between the nominal (observed) data and the real (latent) variable. We estimate all parameters of our models by minimizing the square-error between the integrated variables and their real values, according to regular identification methodologies \cite{Bard1974,Brauer2019}. We proceed by following an hierarchical procedure as done in \cite{bastos2020}.

Our identification procedure also includes a limit\footnote{If more restrictive bounds are used, we mention them  explicitly.} to the parameter values, as follows: $\beta, \beta_S, \beta_A \in [1/20, 2]$, $\gamma, \gamma_S, \gamma_A \in [1/14, 1/2]$, $\rho, \rho_S \in [0.001, 0.2]$, $\alpha_{On} \in [0, 1]$ and $\psi_{Inf} \in [0.3, 0.7]$. These limits are in accordance with those presented by \cite{werneck2020covid, bastos2020}. 

\subsection{Least Square Procedures}

Firstly, we use the estimation of the infection, transmission and death probability, \(\beta\), \(\gamma\) and \(\rho\), respectively,  of the SIRD model in Eq. (\ref{eq:SIRD}). This is done using the data without any enacted the social distancing policy, as observerd in the country from February 25, 2020 to March 22, 2020; \cite{bastos2020} present a through discussion on this matter. The transmission rate parameter $\gamma$ is then fixed for all the subsequent identification procedures, since this parameter is characteristic of the disease. We refer to the $\beta$ value of this step as $b_1$.

Secondly, we assume that $\gamma_K = 0$ during this first period (no social distancing) and estimate the parameters of the SIRDC model from Eq. (\ref{eq:SIRDC}) by minimizing the following square-error:
\begin{equation}
\begin{array}{cc}
\min_{\beta, \rho, \alpha_{On}, \psi_{inf}} & \frac{1}{2} \left( \sum_{t}{\left[f\left((Z(t)-D(t)) - (\hat{I}(t)+\hat{R}(t))\right)\right]^2 + \left[f\left(D(t) - \hat{D}(t)\right)\right]^2} \right)
\end{array},
\label{eq:SIR_model_params_calculation}
\end{equation}
\noindent where $Z_t$ and $D_t$ represent the data provided by the Ministry of Health of Brazil embedded with uncertainty, as gave Eqs. \eqref{Dtuncertain}-\eqref{Ztuncertain}, and $\hat{I}_t$, $\hat{R}$ $\hat{D}_t$ are estimated parameter values using the SIRD model. Note that we use the nonlinear function $f(z) = \ln{(1+z)}$ to correct the exponential characteristic of the series so that the errors of the last values of the series do not dominate the minimization. We  assume that there are no recovered individuals at the beginning of the series, and also that $\psi_0 = 1$ (no-isolation state). We refer to the parameters values of this step as $\beta = b_2^{(q_I, q_D)}$, $\rho = r_2^{(q_I, q_D)}$, $\alpha_{On} = a_2^{(q_I, q_D)}$ and $\psi_{Inf} = i_2^{(q_I, q_D)}$. We limit $\beta$ in $[(1-\delta_{\beta}) b_1, (1+\delta_{\beta}) b_1]$, with $\delta_{\beta}=0.5$. For the models with uncertainty, $\alpha_{On}$ and $\psi_{Inf}$ are also limited in $[(1-\delta_{\alpha_{On}}) a_2^{(1,1)}, (1+\delta_{\alpha_{On}}) a_2^{(1,1)}]$ and $[(1-\delta_{\psi_{Inf}}) i_2^{(1,1)}, (1+\delta_{\psi_{Inf}}) i_2^{(1,1)}]$, with $\delta_{\alpha_{On}}=0.9$ and $\delta_{\psi_{Inf}}=0.3$, that is, these values under uncertainty are limited by a range defined by the simulation without uncertainty.

Thirdly, we estimate the complete SIRASD model. For such, we assume that asymptomatic infected comprise all individuals without symptoms and also those with mild symptoms (that do not need ICU beds), and that symptomatic infected are individuals with moderate to severe symptoms. This line of thought follows the orientation given by the Brazilian Ministry of Health, that incentives people to only seek medical attention if symptoms are moderate or severe, and to stay home otherwise. Therefore, we suppose that \(\gamma_S=\gamma\) and \(\beta_S=\beta\), which corresponds to the simulation with uncertainty only upon the number of deaths (i.e. $q_D \, \neq \, 1$ and $q_I \, = \, 1)$. Since the uncertainty for the number of infected is also related to the asymptomatic individuals, we use \(p=q_I\). We use this values for the infection probability as constant, so that the initial condition for the asymptomatic infected is $I_{A,0} = I_{S,0} (1 - q_I) / q_I$. In this context, we estimate the parameters \(\beta_A\), \(\gamma_A\), \(\rho_S\), \(\alpha_{On}\) and \(\psi_{Inf}\) in order to minimize the following square-error:

\begin{equation}
\begin{array}{cc}
\min_{\beta_A, \gamma_A, \rho, \alpha_{On}, \psi_{inf}} & \frac{1}{2} \left( \sum_{t}{\left[f\left((Z_t-D_t) - (\hat{I}_{S,t}+\hat{R}_{S,t})\right)\right]^2 + \left[f\left(D_t - \hat{D}_t\right)\right]^2} \right)
\end{array},
\label{eq:SIRASD_model_params_calculation}
\end{equation}
\noindent where $\hat{I}_{S,t}$ and $\hat{D}_t$ represent the values obtained with the SIRASD model.

\subsection{Obtained Models}

The previous identification procedure was realized for a large number of possibilities of uncertainty ($q_D$ and $q_I$). Through the sequel, for simplicity, we consider only three SIRASDC models:
\begin{itemize}
    \item The "Nominal" model, which is tuned for $q_D \, = \, q_I \, = \, 1$;
    \item The "Uncertain $1$" model, which is derived from the identification procedure considering $50 \, \% $ more sub-notified deaths and $30$ times more infected individuals than reported ($q_D = 2/3$ and $q_I = 1/30)$;
    \item And the "Uncertain $2$" model, which is likewise found through the identification procedure for $50 \, \% $ more sub-notified deaths and $15$ times more infected individuals than reported ($q_D = 2/3$ and $q_I = 1/15)$.
\end{itemize}
The respective model parameters, found through the identification procedure detailed in Section \ref{sec:estimation}, are presented in Table \ref{ModelParametersTable}.

\begin{table}[htbp]
    \caption{\label{ModelParametersTable} Control Simulation: Considered Model Parameters.}
    \centering
	\begin{tabular}{|c | c c | c c | c c|} 
		\hline \hline
        Model & $\beta_A$ & $\gamma_A$ & $\beta_S$ & $\gamma_S$ & $p$ & $\rho$ \\ \hline
        Nominal & $0.44$ & $0.1272$ & $0.4230$ & $0.0695$ & $0.016$ & $0.049$ \\ \hline
        Uncertain 1 & $0.3690$ & $0.0952$ & $0.4307$ & $0.1395$ & $ 0.066$ & $0.1462$ \\
        Uncertain 2 & $0.4307$ & $0.1395$ & $0.3723$ & $0.0985$ & $0.0322$ & $0.1461$ \\ \hline
        \hline
    \end{tabular}
\end{table}

\subsection{Some results}

\begin{figure}[htb]
	\centering
   \includegraphics[width=\linewidth]{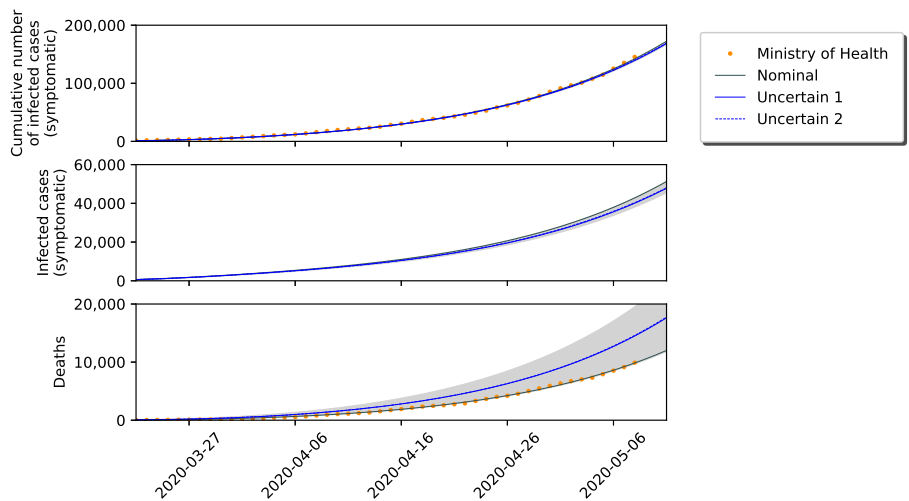}
	\caption{Short-term simulation for the SIRASD model: Nominal vs. Uncertain}
	\label{UncvsNom_short}
\end{figure}

To conclude this Section, we show some short-term simulation results in Figure \ref{UncvsNom_short}. This Figure represents the SIRASDC model running with respect to known data, from 20/03/27 until 20/05/08. The differences between the speed of the COVID-19 spread considering nominal and uncertain conditions are considerable. 

This Figure depicts the differences between the models and also marks how the nominal model perfectly fits the dataset from the Ministry of Healthy. 

Clearly, the amount of sub-notification plays a significant role in the analysis of this pandemic. If one does not take the uncertainty into account, wrong and rash decisions may be performed. We must stress that an increase on the number of infected individuals (larger sub-notification) means that the mortality rate of the disease decreases, while it increases for a larger sub-notification margin with respect to the number of deaths.

\section{The Predictive Control Strategy}
\label{sec:control}

In this Section, we develop the second main contribution of this
  work, i.e. the optimal control strategy aiming to mitigate the impacts of the COVID-19 pandemic. In practice, the resulting control law which should be implemented by the means of social distancing policies, conducted through orientation by the local government.  
  
  For this goal, we consider the control-appended models (SIRDC and SIRASDC), which are now regulated under a closed-loop scheme. In fact, the proposed control strategy is formulated with respect to SIRASDC model, as explained in the sequel.
  
  We must consider that the predictive control strategy is to be synthesized based on a model that \textit{a priori} encompasses the dataset mismatches (uncertainty in terms of sub-notification). Taking into account a model that considered that all measures (deaths, infections) are not exact can further improve the outcome of the control policy, since it will be more conservative with respect to social isolation measures. In some sense, this kinds of strategy is a robust MPC feedback procedure, because it is based on a worst-case pandemic level. The application of the MPC feedback without the uncertain realization of the model can lead to performances which may borderline the use of the available ICU beds in the country, i.e. the nominal prediction model may give a result in terms of infections which is less than what is observed in practice. This situation would lead to necessity of possibly longer periods of social isolation. Then, such robust MPC procedure is able to avoid this kind of situation and it can be able to significantly reduce the number of fatalities.

\subsection{Control Objectives}
\label{ControlObjectivesSec}
As previously discussed, social isolation measures are necessary since they are able to "flatten" the COVID-19 spread curve. Figure \ref{IsolationNeeded} illustrates this issue, which shows a simulation for the evolution of symptomatic individuals $I_s(t)$ over time with no social isolation $\psi (t) = 1$ and hard social isolation ($\psi = 0$, which represents a complete lock-down situation). Clearly, when social isolation policies are implied (mathematically, for $\psi < 1$), the infection peak is postponed and reduced.

Therefore, the control objectives regarding COVID-19 isolation policies are the following:
\begin{itemize}
\item To reduce the peak of symptomatic individuals, as much as possible;
\item To ensure that the peak is smaller than the number of available ICU beds (a tolerance factor can be included);
\item To determine isolation policies for as little time as possible (in order to mitigate the effects of isolation on economy);
\item And to avoid shifting between states (isolation or not), maintaining a minimal period of $N_m$ days in each condition.
\end{itemize}

These objectives can be mathematically expressed, respectively, as follows:
\begin{enumerate}[label =(\alph*)]
\item By minimizing the amount of symptomatic individuals $I_S$;
\item By ensuring that $I_S \, \leq \, n_{ICU}(1 + \xi)$ for all $t$, where $n_{ICU}$ represents the amount of ICU beds and $\xi \, \in (0 \, , 1)$ a tolerance factor (which should also be minimized);
\item By minimizing the isolation policy $u(t)$, which should be $0$ (no isolation) for as long as possible;
\item To ensure that $u(t)$ is piece-wise constant and maintained at each state ($0$ or $1$) for $N_p$ samples, where $k =\frac{t}{T_s}$ stands for the sampling unit and $T_s$ the sampling period. We denote this as a dwell-time constraint on $u$.
\end{enumerate}

\subsection{Control Law}
\label{ControlLawSec}
As remarked by \cite{silveira2005piecewise}, the implementation of any feedback control policy derived from $u(t)$ to real biological system may light upon two complicating issues:
\begin{enumerate}
\item Feedback control in the continuous time-domain $t$ would require the measurement of the involved variables at every instant of time, which is obviously not possible. Anyhow, since the COVID-19 spread has "slow" dynamics, discretized versions of the SIRDC / SIRASDC continuous models can be used. Given that new measurements are available each day, the discrete sampling is of $T_s \, = \, 1 \rm{day}$; 
\item The control signal should, essentially, model the human action on the studied ecosystem. As previously argued, the human correspondence to the isolation policies have already been included to the SIRDC and SIRASDC models, through the dynamics of $\psi(t)$.
\end{enumerate}

\begin{remark}
\label{EulerDiscretizedRemark}
The discretization method used in this paper is very usual: all derivative functions $\frac{dx}{dt}(t)$ are approximated by the deviance along on sampling period, this is: $\frac{dx}{dt}(t) \approx \frac{\left(x(k+1) - x(k) \right)}{T_s}$. This is usually referred to as Euler/forward discretization \cite{skogestad2007multivariable}.
\end{remark}

Henceforth, this paper considers a piece-wise constant control signal $u(t)$, generated from periodic measurements, every $T_s = 1$ day. Considering control increments denoted $\Delta u(k)$, it is implied that:
\begin{eqnarray}
\label{Piecewiseu}
u(k)&=& u(k-1) +  \Delta u(k) \, \text{.}
\end{eqnarray}

The actual control law that the controller applies to the system is $u(t) \, = \,u (kT_s)$ for the whole time during each sampling period interval, i.e. $\forall t \, | \, kT_s \, \leq t \, \leq \, (k+1)T_s$. This is clearly a piece-wise constant signal due to Eq. \eqref{Piecewiseu}. 

Apart from being piece-wise constant, the control law $u$ must obey another main restriction, in order to be implementable in practice: it must depart from $u(0) = 1$, which is the last observed social distancing policy (the country is still in an isolation condition). It must be noted that $t = 0$ stands for the instant corresponding to the last sampled field data (infected, deaths dating $29$ April).

\subsection{The Optimal On-Off Control Framework}

To design an optimal controller which determines the On-Off control policy $u(t)$ according to the previous discussion, we will follow the Model Predictive Control formalism.

Sometimes named moving/sliding horizon control, the MPC concept is quite straightforward for the optimal control of constrained system. The basic MPC formulation resides in the solution of an optimization problem with respect to a sequence of control actions $U_k$, at each discrete instant. This optimization is written in terms of a process prediction model, performance goals and constraints, which are handled explicitly.


Therefore, we consider an well-posed quadratic function $J$ which is minimized seeking to accomplish objectives (a) and (c) (reducing infections and reducing the social isolation periods) presented in Section \ref{ControlObjectivesSec}. This cost function is analytically expressed as:
\begin{eqnarray}
\label{JcostMPC}
J&=& \sum_{j = 1}^{N_p} \frac{\left(I_S(k+j|k)^TQ_I I_S(k+j|k)\right)}{(I_S^{max})^2}  + \sum_{j = 0}^{N_p -1} u(k+j|k)^TQ_u u(k+j|k) \, \text{,}
\end{eqnarray}
where $N_p$ is a given prediction horizon, and $Q_u$ and $Q_I$ are weighting matrices. The notation $(k+j|k)$
stands for a model-based prediction for instant $k + j$
made at instant $k$. The constant $I_S^{max}$ stands for the maximal possible value of $I_S$, with respect to open-loop simulations (as in Figure \ref{IsolationNeeded}); we include this constant to ensure that the magnitude of the first and second term of $J$ are the same (i.e. normalization).

The vector of control efforts inside
                     the prediction horizon $U_k$ (to be optimized) is also presented:
\begin{eqnarray}
\label{BigU}
U_k &=& \left[\begin{array}{cccc} u(k | k) & u(k + 1 | k) &
                                                                     \dots
                         & u(k + N_p -1 |k)\end{array}\right]^T \, \text{.}
\end{eqnarray}
From this control sequence, at each sampling instant $k$, one takes the first entry $u(k)$ and applies the control signal according to Eq. \eqref{Piecewiseu} to the controlled COVID-19 spread process (SIRDC / SIRASDC models).

Notice that if one simply minimized the previous cost function $J$, at each sampling instant $k$, with respect to a control sequence $U_k$, the results would be a control sequence $U_k$ which provides a trade-off (according to weights $Q_I$ and $Q_u$) between the minimization of infected individuals and control effort ($u$ gets closer to $0$ and, thus, less isolation is implied).

Anyhow, for an appropriate application of this paradigm, the constraints of each $u(k+j|k)$, as given in Section \ref{ControlLawSec}, should be taken into account by the optimization procedure. It follows that:
\begin{eqnarray}\label{Constraint1}
I_S(k+j|k) &\leq & n_{ICU}(1 + \xi) \quad \forall j \, = \, 1,\dots , N_p \, \text{,} \\\label{Constraint2}
0 \,\, \leq \, \, u(k+j|k) &\leq & 1 \quad \forall j \, = \, 0,\dots , N_p-1 \, \text{,} \\\label{Constraint3}
u(k+j+1|k) &= & u(k+j|k) + \Delta u (k+j|k) \quad \forall j \, = \, 0,\dots , N_p-1 \, \text{,} \\\label{Constraint4}
 \Delta u (k+j|k)  &\text{is} & \text{binary} \quad \forall j \, = \, 0,\dots , N_p-1 \, \text{,} \\ \label{Constraint5}
 \Delta u (k+m|k) &= & 0 \text{ if } \left\|\Delta u (k+j+1|k) - \Delta u (k+j|k)\right\| = 1 \quad \forall m \, = \, j,\dots , j+N_{m} \, \text{.} 
\end{eqnarray}

Notice that: 
\begin{itemize}
\item Eq. \eqref{Constraint1} ensures that the peak of infections is reduced and does not surpass $ n_{ICU}(1 + \xi)$;
\item Eq. \eqref{Constraint2} ensures the control signal is bounded within the social isolation limits, Eq. \eqref{Constraint3} ensures that this law is piece-wise constant and Eq. \eqref{Constraint4} implied that the variation is binary (so that the control is, in fact, "On"/"Off");
\item Eq. \eqref{Constraint5} implies that a minimal dwell-time of $N_{m}$ samples must be accounted for, i.e. the control determines that $u$ stays at a given state "On" or "Off" for a minimal period of $N_{m}$ days. It is implied that $N_p > N_m$.
\end{itemize}

Notice that the slack/tolerance variable $\xi$ defined in in Section \ref{ControlObjectivesSec} should also be minimized, which means that a $\xi ^T Q_I \xi$ is included to the cost function $J$, as follows: 
\begin{eqnarray}
J &=& \sum_{j = 1}^{N_p} \frac{\left(I_S(k+j|k)^TQ_I I_S(k+j|k)\right)}{(I_S^{max})^2}  + \sum_{j = 0}^{N_p -1} u(k+j|k)^TQ_u u(k+j|k) + \xi^T Q_I \xi\, \text{.}
\end{eqnarray}

Therefore, bearing in mind this previous discussion, the MPC approach to mitigate the effect of COVID-19 spread consists in minimizing the cost function $J$ at every discrete-time step $k$, with respect to the previously discussed constraints and taking into account a discretized version of the SIRASDC model, with $T_s \, = \ 1$ day. One can mathematically express this problem as follows:
\begin{eqnarray}
\label{MPCProblemFinal}
\min_{U_k} &&  J
\end{eqnarray}
\begin{eqnarray}
\nonumber
\begin{array}{cc}
\text{subject to:} & \left\{\begin{array}{c}
\text{Discrete SIRASDC Model, Eqs. \eqref{Eqdiscrete1}-\eqref{Eqdiscretefinal}} \\
\text{Peak reduction constraint: Eq. \eqref{Constraint1}} \\
\text{Control signal constraints: Eqs. \eqref{Constraint2}-\eqref{Constraint5}}
 \end{array} \right.
\end{array}
\end{eqnarray}

Note that the discrete SIRASDC model is found by applying the discretization method detailed in Remark \ref{EulerDiscretizedRemark} to Eq. \eqref{eq:SIRASDC}. This discretized model is then extended as a prediction model, describing future instants $(k+j)$ with respect to the information at $(k)$, this is:
\begin{eqnarray}\label{Eqdiscrete1}
S(k+j+1) &=& S(k+j) -T_s\psi(k+j)\left(\beta_A I_A(k+j) + \beta_S I_S(k+j)\right)\frac{S(k+j)}{N(k+j)} \, \text{,} \\
I_A(k+j+1) &=& I_A(k+j) - T_s\gamma_A I_A(k+j) \\ \nonumber &+& T_s (1-p) \psi(k+j) \left(\beta_A I_A(k+j) + \beta_S I_S(k+j)\right)\frac{S(k+j)}{N(k+j)} \, \text{,}\\
I_S(k+j+1) &=& I_S(k+j) - T_s\frac{\gamma_S I_S(k+j)}{1-\rho}   \\ \nonumber &+& T_s p \psi(k+j)\left(\beta_A I_A(k+j) + \beta_S I_S(k+j)\right)\frac{S(k+j)}{N(k+j)} \, \text{,} \\
D(k+j+1) &=& D(k+j) + T_s\frac{\rho}{1-\rho}\gamma_S I_S(k+j) \, \text{,} \\
\psi (k+j+1) &=& \psi(k+j) + T_s\alpha_{\mathrm{Off}}(1-\psi(k+j))(1-u(k+j)) \\ \nonumber &+& T_s\alpha_{\mathrm{On}}\left(K_\psi(k+j)\psi_{\mathrm{inf}} - \psi(k+j)\right)u(k+j) \, \text{,} \\ \label{Eqdiscretefinal}
K_\psi(k+j) &=& 1- \gamma_K \frac{\rho}{1-\rho}\frac{I_A(k+j)}{N(k+j} \, \text{.}
\end{eqnarray}

\section{Simulation Results, Forecasts and Discussion}
\label{sec:discussion}

In this Section, we present simulation forecasts using the SIRDC/SIRASDC with parameters identified in Section \ref{sec:estimation}. The following results were obtained with the aid of Matlab software, Yalmip toolbox and BNB solver.

In the sequel, the baseline threshold in the $I_S$ curves represent the number of available ICU beds in the country (see Figure \ref{ICUbedsBr}). The maximal threshold stands for an incremented number of ICUs (twice the baseline value), accounting for field hospitals and emergency ICUs that have been made specifically for the COVID-19 pandemic.

The following control results were obtained considering the SIRASD models (Nominal, Uncertain 1 and Uncertain 2), as presented in Section \ref{sec:estimation}.

Before presenting the actual results, we must affirm that the forecasts and arguments that we present in the sequel should not understood by the reader as incontrovertible truths. These forecast are model-based simulations which depend on a number of factors and initial conditions. Furthermore, we must stress that we have aggregated the whole set of Brazilian data in order to provide a general view of the country. However, if anyone intends to use the proposed method to help the formulation of public health policies, we suggest its application to datasets of smaller regions, that share the same hospital chain. We note, as illustrated in Figure \ref{ICUbedsBr}, that different regions of the country are facing different levels of the pandemic.

As evidenced in Section \ref{sec:estimation}, the used models grasp the behaviors the SARS-CoV-2 virus dynamics quite accurately, but this does not means that the future predictions are unmistakable. For a fact, we cannot ensure that the social isolation measures will be strictly followed by the population, as we cannot ensure that other factors may come to help ease the spread of the disease (such as vaccines). What we mean by this is that the goal of this work is guide public policies regarding social isolation specially by taking into account the role of uncertainty and sub-notification.  

Due to the fact that the proposed model cannot exactly predict the pandemic dynamics, to apply some control polcy conceived based on a nominal model may lack conservatism. This could lead to catastrophic results, risking high levels of mortality. Any possible control policy that the government implements through social distancing measures must be based on recurrent (worst-case) model parameter estimations and recalculations of the optimization problem. One cannot use the models derived with the parameters presented in Table \ref{ModelParametersTable} as if they would not change along time. The correct measure is to take into account uncertainty-embedded models, performing the identification procedure detailed in Section \ref{sec:estimation} every day (when new datasets are available). Such adaptive control procedure (with model tuning and model-based control optimization) would be much more prudent, requiring constant measuring, monitoring, parameter estimation and control computations. As discussed by \cite{kohler2020robust}, feedback is utterly necessary to ensure a reliable handling of the SARS-CoV-2 outbreak. This is especially critical in Brazil, due to the high level of uncertainty on the datasets.

Through the sequel, the dashed lines represent the results with uncertainty (solid dash, Uncertain 1 model; dot dash, Uncertain 2 model), while the solid lines account for nominal conditions.  

\begin{figure}[htb]
	\centering
		\includegraphics[width=\linewidth]{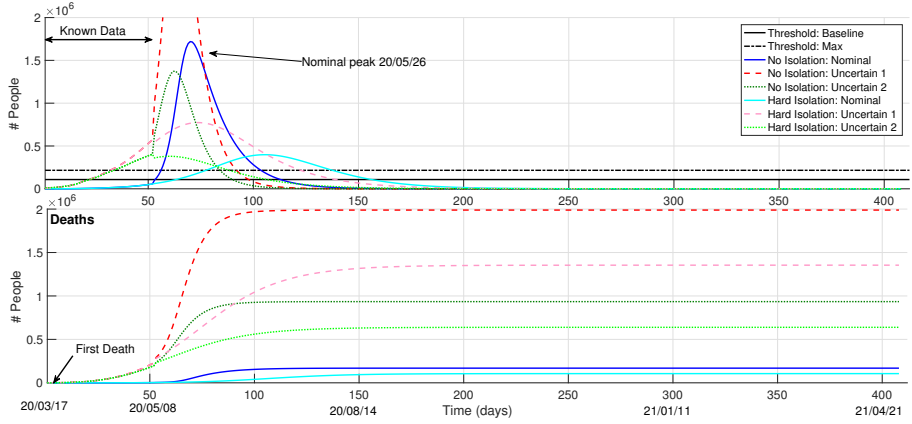}%
	\caption{Necessity of Social Isolation 2: Model-based Forecasts}
	\label{IsolationNeeded2}
\end{figure}

Considering these models, Figure \ref{IsolationNeeded2} shows the simulation for roughly one year dating from the last data sample ($20/03/17$), considering a total lock-down condition ($u = 1$) and a no-isolation ($u = 0$) case. The first 52 samples represent the known dataset, whereas the following samples stand for predicted data. Clearly, even if a hard isolation is enacted, the Brazilian health system will still face issues with large amounts of COVID-19 patients, with a nominal peak forecast to $26^{\text{th}}$ of May. The nominal collapse of the healthy system (threshold) dates very soon, May 23. The amount of deaths expected with the uncertain model is unprecedented. Of course, each life matters and $2$ million deceased individuals is a lot to bare. Psychological and social traumas will mark the country. A hard isolation could be able to save more than one million lives, taking into account the results achieved with the worst-case uncertainty scenario.

With respect to the forecasts presented in Figure \ref{IsolationNeeded2}, we must also stress that the possibility of herd immunity must be discarded. These results corroborate the conclusions presented by \cite{kohler2020robust}, which indicate that neither a complete eradication of the virus nor herd immunity are possible options to attenuate the COVID-19 pandemics without the availability of a vaccine. These results also go along the lines of \cite{hellewell2020feasibility}.

Considering control results, the MPC optimization procedure from Eq. \eqref{MPCProblemFinal} is solved for different cases of $N_m$ (minimal amount of days in each state: isolation, no isolation). For such, the weighting matrices $Q_I$ and $Q_u$ are taken to imply an adequate trade-off between peak reduction and social isolation. Since the occupancy rate of ICU beds in the country dating $20/05/08$ is considerably high, we chose $Q_I = 0.9$ and $Q_u = 0.1$, which means that the MPC makes "more effort" to reduce the amount of infected individuals then to restore a no-isolation policy, which is reasonable considering the observed situation. The control horizon is fixed as $N_p = 60$ days (the MPC makes predictions for two-months ahead of each sampled $k$, day).

\begin{figure}[htb]
	\centering
		\includegraphics[width=\linewidth]{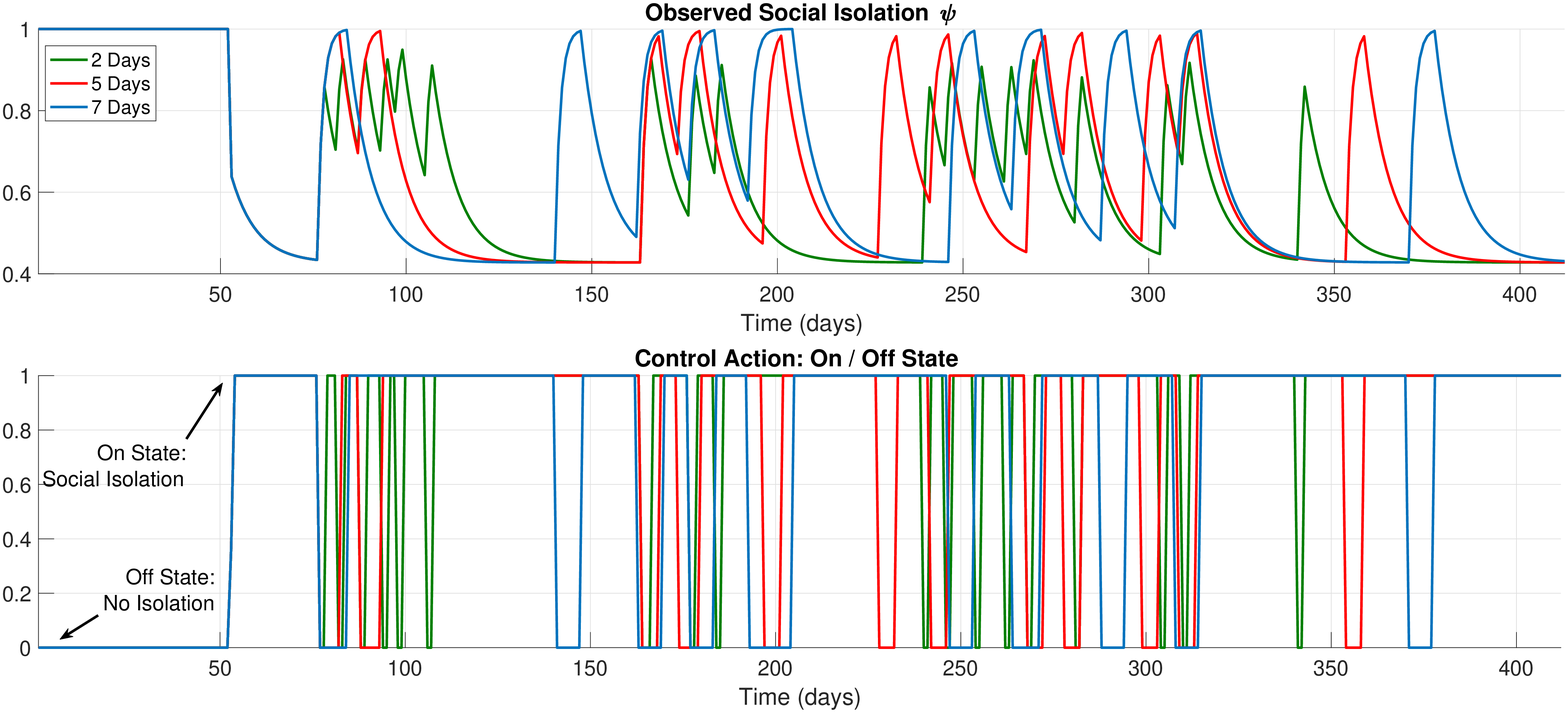}%
	\caption{Control Policy, $N_m = 2$, $5$ and $7$ Days, Excessive Shifting}
	\label{ControlPoliciesTimin2and5}
\end{figure}

\begin{figure}[htb]
	\centering
		\includegraphics[width=\linewidth]{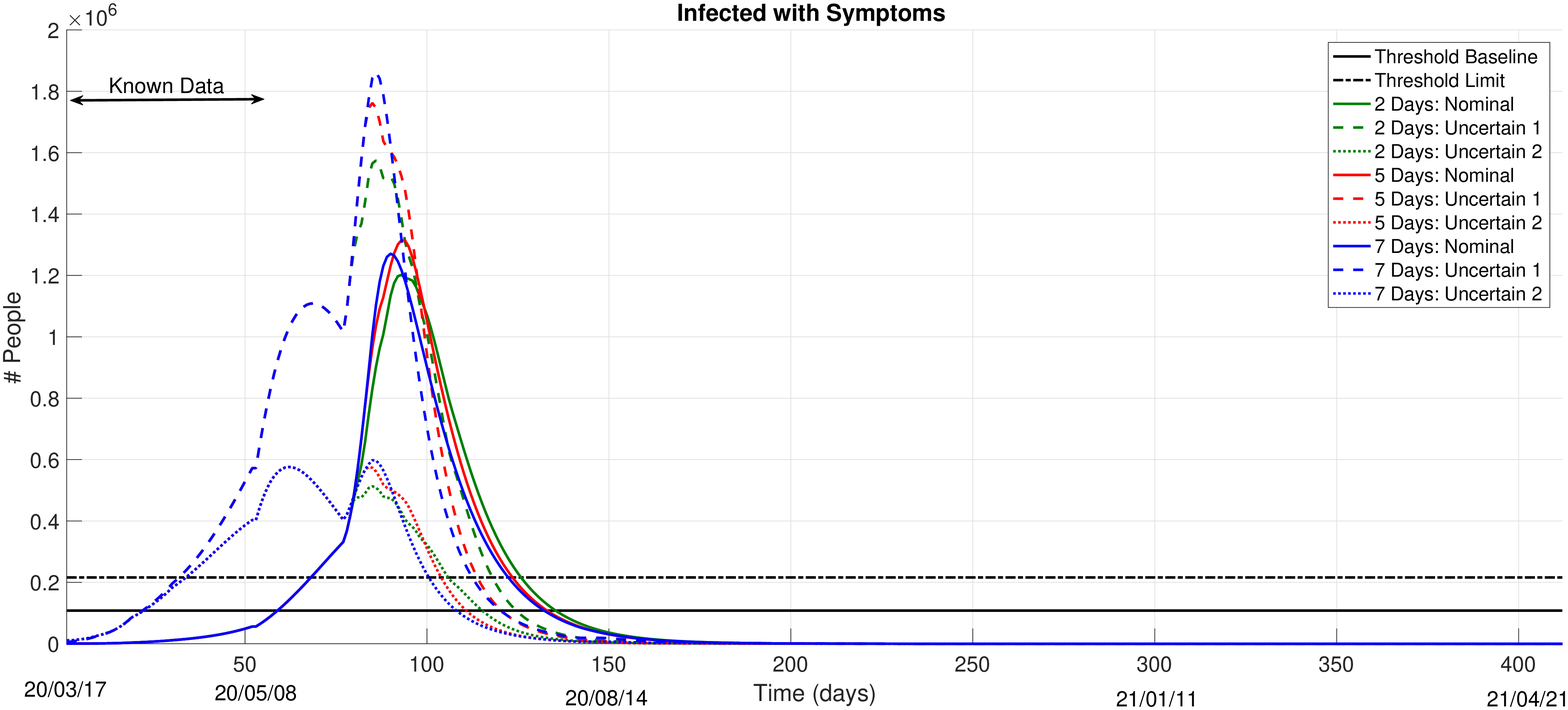}%
	\caption{$N_m = 2$, $5$ and $7$ days, Resulting regulation}
	\label{SymptomaticTmin2and5}
\end{figure}

Firstly, we show the results for $N_m = 2$, $5$ and $7$ days of the minimal days in each state condition. The decision by the MPC optimization and the resulting enacted social isolation measure ($\psi$) are shown in Figure \ref{ControlPoliciesTimin2and5}. The resulting effects on the amount of symptomatic individuals is shown in Figure \ref{SymptomaticTmin2and5}. We must state that, for $N_m \leq 7$ days, the amount of shifting in the observed social isolation variable is quite intense. Furthermore, the obtained results with these values for $N_m$ were not enough to reduce the infection peak, as observed. Note that this kind of policy would hardly ever be implementable, since a confusing message would be passed to the society. The frequent changing between isolation or no-isolation would not be strictly followed, which is certainly unwanted. Therefore, these results are not considered as practical or viable.

\FloatBarrier

\begin{figure}[htb]
	\centering
		\includegraphics[width=\linewidth]{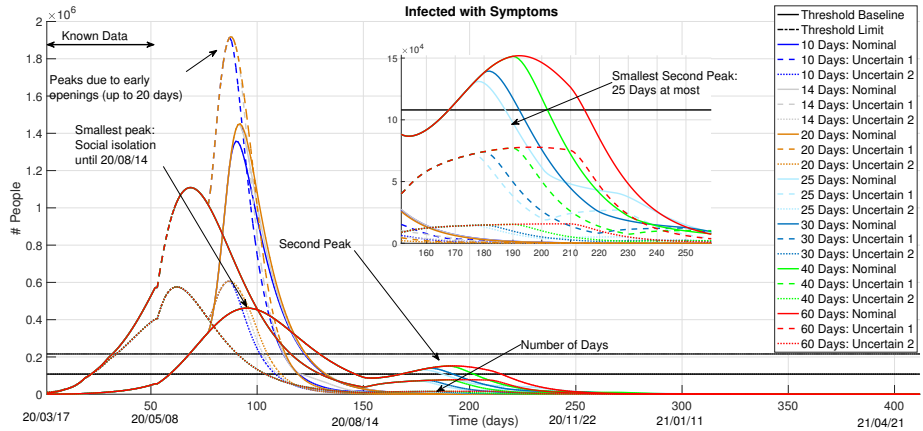}%
	\caption{$N_m = 10,14,20,25,30,40$ and $60$ Days: Infected with Symptoms}
	\label{Symptomatic7to60}
\end{figure}

\begin{figure}[htb]
	\centering
		\includegraphics[width=\linewidth]{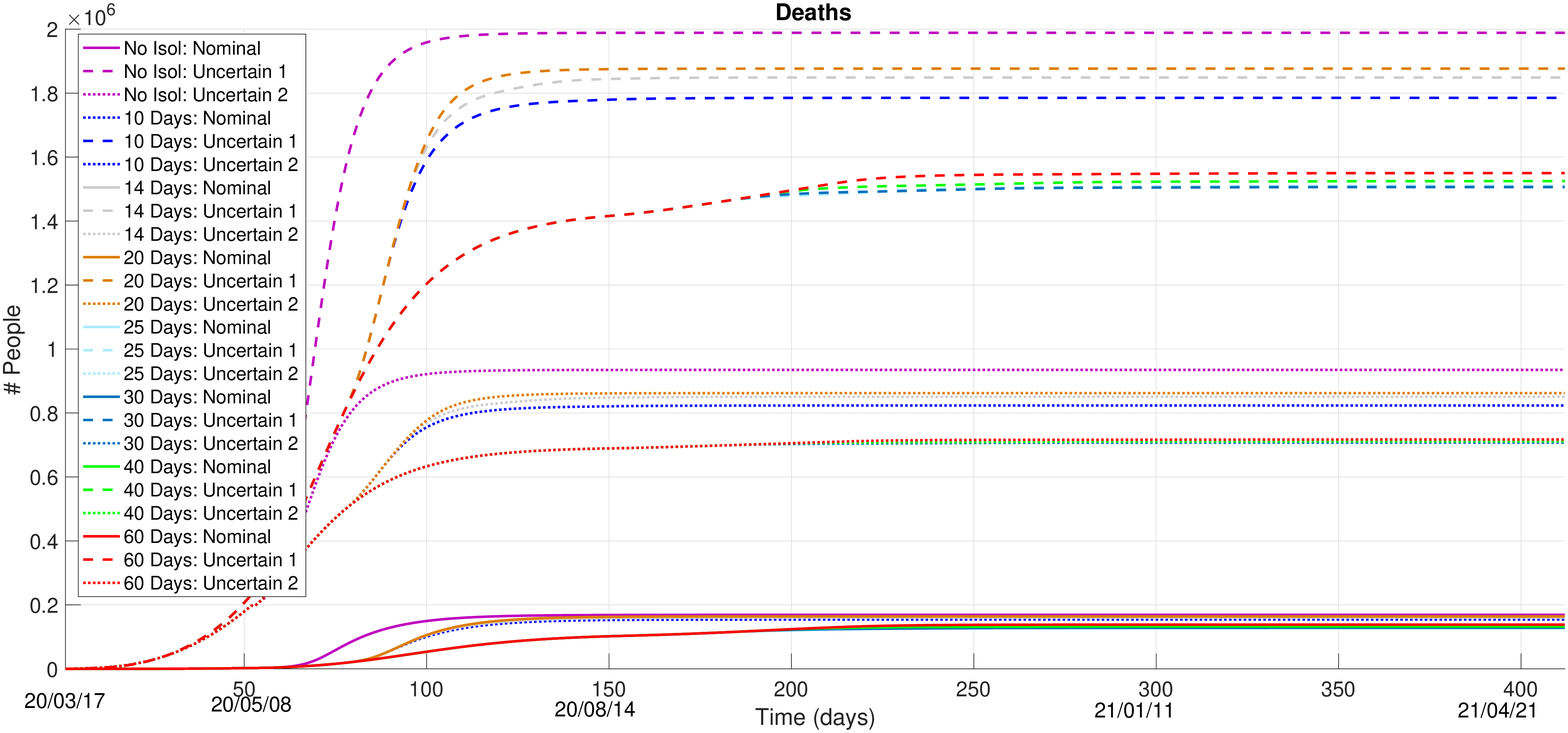}%
	\caption{$N_m = 10,14,20,25,30,40$ and $60$ Days: Deaths}
	\label{Deaths7to60}
\end{figure}

Then, Figure \ref{Symptomatic7to60} presents the obtained control results for $N_p = 10, 14, 20, 25, 30, 40$ and $60$ days, regarding the $I_S$ curve. This Figure contains a lot of information, which we try to explain by parts:
\begin{itemize}
    \item If social isolation is not maintained until roughly August 14, even the shortest "openings" (days in reduced isolation policies) could be catastrophic. Any possible reduction of the hard social isolation measures, before this date, would result in an infection peak which would surpass the amount of available ICU beds in the country in over seven times (considering the worst-case uncertainty). This is very significant and thus, any possible social isolation reduction should not proceed before the initial infection curve starts to decay (in both nominal and uncertain conditions);
    \item Therefore, smallest peaks of infections occurs if the isolation measure is kept at least until August 14. After this date, when relaxations in these measures are enacted, a second infection peak will certainly appear. This second peak is due to the fact that social isolation is reduced by the MPC law after the decay of $I_S$ running from first peak (August 14). This second peak dates roughly October 3.
    \item The second peak of infection is reduced with a smaller number of days in a no-isolation mode. This means that, after August 14, the MPC control action which results in the smallest values for $I_S$ are those with, at most, periods of $25$ days in the no-isolation mode. The smallest amount of "open" periods,  better the results, as expected. 
    \item Note that as $N_m$ increases, this second peak of infection also increases because the amount of minimal days determined for a no-isolation (or reduced-isolation) policy forces a peak increase, which is later treated by a total isolation after its decay. The amount of deaths are given in Figure \ref{Deaths7to60}. Depending on the amount of days in a no-isolation condition, the amount of deaths may range from $0.13$ to $1.88$ million individuals, according to the uncertain models. 
\end{itemize}

\begin{figure}[htb]
	\centering
		\includegraphics[width=\linewidth]{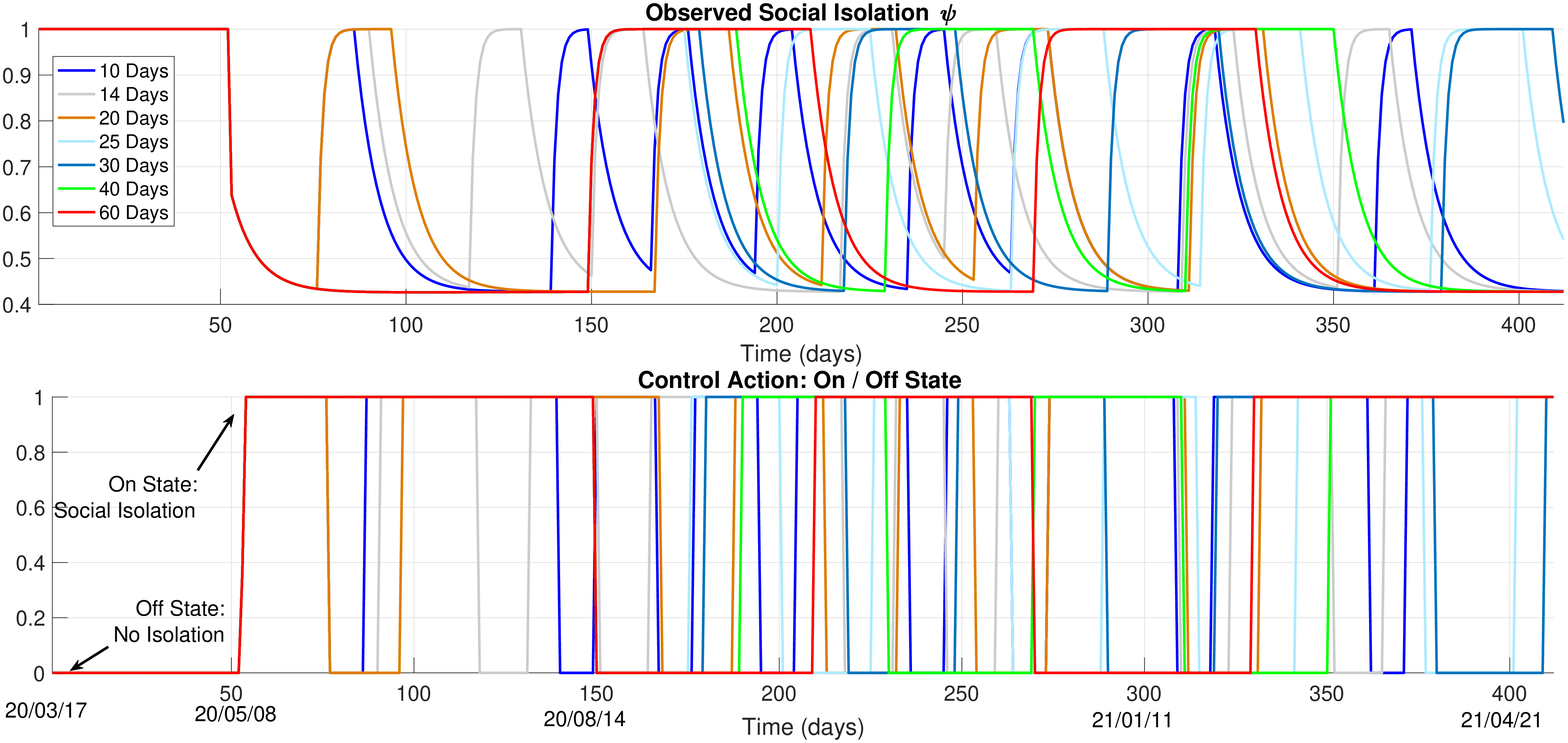}%
	\caption{Control Policy: $N_m = 10, 14, 20, 25, 30, 40$ and $60$ Days of Social Isolation}
	\label{ControlPolicies25to60}
\end{figure}

Regarding these obtained results, it seems reasonable to us to ponder the following issues:
\begin{itemize}
    \item The uncertain model forecasts quite harsh infection scenarios. Even though the considered uncertainty is quite high ($15, 30$ times more cases), it offers us a worst-case forecast to determine public policies. With such uncertain model in mind, it seems evident why social distancing measures are so important right now and why they should not be dropped, despite their possible economical side-effects. It seems extremely necessary for public policies to offer alternative solutions to those without jobs or economically suffering due to the social isolation. 
    \item No social distancing measures should be relaxed before mid-August ($20/08/14$). This would definitely help in avoiding the collapse of the Brazilian health system.
    \item If social distancing is to be relaxed, this should not be done before the first infection peak starts to decay (beginning of August) and the no-isolation periods should be the minimal amount of days possible. To ensure heath safety, a conservative measure indicates that such paradigm of recurrent short periods of reduced isolation, followed by hard isolation periods would proceed until roughly $2021/01/21$. This paradigm would be helpful to ensure that the SARS-CoV-2 virus does not cause further infection peaks and to mitigate the amount of deaths.
\end{itemize}

The resulting control policies from the MPC procedure, for the different values for $N_m$, are shown in Figure \ref{ControlPolicies25to60}. These curves indicate, roughly, when to determine social isolation measures and when to set them off. In fact, the actual implemented policy would depend on a daily update of the MPC results with measured datasets. Anyhow, these results indicate a forecast of roughly when to determine or call off these measures. The best result, in terms of infections, would be to follow the Social Isolation state until mid-August, an then relax this measure with small periods (that should definitely not surpass $25$ days).

It seems, mathematically speaking, that even if the SIRASD model has a new degree-of-freedom (which is the decision variable $u$, to determine when to determine social isolation), the resulting optimization points out that the best option is to maintain isolation for as long as possible. Even if allowing social contact for a while, the optimization finds minima solutions of $J$ for the smallest number of days with contact and, then, once again determines isolation. 

The COVID-19 is quite worrisome and presents devastating social and economic effects. Biology literature points out that social isolation is necessary. Using mathematical models and optimization, the answer is the same.

\FloatBarrier   

\section{Conclusions}
In this paper, we investigate an optimization-based solution for social isolation measures of the COVID-19 spread for the Brazilian context. Since recent works have warned against the large order of sub-notification in Brazil, we take uncertainty into account to determine nominal and uncertainty dynamic models of the COVID-19 pandemics. Such uncertainty-embedded models are SIR-kind equations which also consider a new variable, which accounts for the average response of the population to social distancing measures (as determined by the government). A robust Model Predictive Control framework is designed for the regulation of the COVID-19 through the means of such social isolation policies. The MPC is derived as an optimal On-Off social distancing planner.

In this paper, we have tried to expose some essential insights regarding sub-notification and how possible relaxations of social distancing can be performed in the future. Below, we summarize the main findings of this paper, enlightening the key points:
\begin{itemize}
    \item The presented results corroborate the hypothesis formulated in \cite{hellewell2020feasibility} and also discussed in \cite{THELANCET20201461}, with respect to the Brazilian scenario: herd immunity cannot be considered a plausible solution, offering great risk and leading to elevated fatality. Furthermore, as illustrate \cite{silva2020bayesian, rocha2020expected,rodriguez2020covid}, vertical isolation is also not an option for the time being, since we do not have the means to formulate an efficient public policy to separate the population at risk from those with reduced risk, due to multiple social-economical issues of the country.
    \item Since the spread of the SARS-CoV-2 virus is inherently complex and varies according to multiple factors (some which are possibly unmodelled and external), exact prediction of the pandemic dynamics is not possible. Therefore, the correct control procedure should be based on a recurrent (daily) model tuning and re-calculation of the control law, always taking into account the uncertainty margins.
    \item The simulation forecasts found through the MPC optimization procedure, which accounts for the uncertainty in the spread of the disease, indicate that, at least for now, only one answer is available: maintain social isolation for as long as possible, without relaxing it before mid August 2020. This is a rather strict suggestion, but seems to be the sole possible way to attenuate the (already high) levels of the virus in Brazil. The forecasts also indicate a prediction for the infection peak in the country dating very soon, May 26, with a second (and larger) peak possibly arising in October. The control policy, in terms of social isolation, shows that relaxations (loosening the isolation measures) should be performed in, at most, periods of $25$ days of reduced-isolation, after the first infection peaks has passed, until roughly January 2021.
\end{itemize}

Synthetically, we must stress that this paper presents only qualitative results of how an optimization-based On-Off strategy can be formulated regarding the COVID-19 spread, regarding the Brazilian context. Since the country as been experiencing an unwillingness to formally start harder social isolation measures \cite{THELANCET20201461}, the social and economic costs of the pandemic might be brutal. The Authors hope that the proposition herein formalised can serve to help determining adequate public health policies from now on.

\section*{Acknowledgment}
 D. O. C. and J. E. N. acknowledge the financial support of CNPQ under respective grants $302629/2019-0$ and $304032/2019-0$.
 
\subsection*{Notes}
The authors report no financial disclosure nor any potential conflict of interests.


\bibliographystyle{model5-names}
\bibliography{references}

\end{document}